# Combination of Subtractive Clustering and Radial Basis Function in Speaker Identification

*Ibrahim A. Albidewi, Yap Teck Ann*

*Faculty of Computing and Information Technology, King Abdulaziz University, Jeddah Saudi Arab*
*Faculty of Computer Science and Information Systems, University Teknologi Malaysia, 81310 Skudai, Johor, Malaysia*

**Abstract**— Speaker identification is the process of determining which registered speaker provides a given utterance. Speaker identification required to make a claim on the identity of speaker from the Ns trained speaker in its user database. In this study, we propose the combination of clustering algorithm and the classification technique – subtractive and Radial Basis Function (RBF). The proposed technique is chosen because RBF is a simpler network structures and faster learning algorithm. RBF finds the input to output map using the local approximators which will combine the linear of the approximators and cause the linear combiner have few weights. Besides that, RBF neural network model using subtractive clustering algorithm for selecting the hidden node centers, which can achieve faster training speed. In the meantime, the RBF network was trained with a regularization term so as to minimize the variances of the nodes in the hidden layer and perform more accurate prediction.

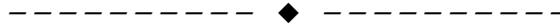

## 1 INTRODUCTION

Pattern recognition is the scientific discipline whose goal is the classification of objects into a number of categories or classes[1]. Pattern recognition in speech can be divided into speaker recognition and speech recognition. The different between speaker recognition and speech recognition are speaker recognition is recognizing who is speaking but speech recognition is recognizing what is being said.

Speaker recognition technologies have two major applications that are speaker identification and speaker verification. The goal of speaker identification is to recognize the unknown speaker from a set of N known speakers. On the other hand, the goal of speaker verification is evaluate whether the claimed identity is correct or not when the unknown speaker presents a speech sample.
The performance of the speaker recognition system is dependent on few factors. Speech variations are hard to distinguish even the same particular effect is observed on the speech signal. The speech variation can divided into six categories - intra-speaker variations, inter-speaker variations, model size, robustness, modeling and accuracy.

One of the factors that will influence the performance of the speaker recognition system is amount of data. The population of data complexity increase the of degradation rate inter-speaker variation. Beside this, the ways to build speaker models using statistical methods in data training might influence system decision biased the verification environment is different from the enrollment[2].

A static model is proposed which employed as an input pattern of Multilayer Perceptron (MLP) network [24]. A MLP network consisting input layers, hidden layers and output layer. The network has a simple interpretation as a form of input-output model, with the weights and thresholds (biases) the free parameters of the model. Such networks can model functions of almost arbitrary complexity, with the number of layers, and the number of units in each layer, determining the function complexity[23]. However, MLP network has some disadvantages. First, MLP needs input pattern of fixed length[24] which will cause the difficulties when dealing with the speech data. Another disadvantage are MLP increasing the number of connections in the network. The increasing of the number of connection will increases the training time and make it more probable fall into a poor local minima.

## 2 IDENTIFICATION TAXONOMY

Speaker recognition can be divided into speaker verification and speaker identification. Speaker verification is speaker claims to be of a certain identity and the voice is used to verify this claim. On the other hand, identification is the task of determining an unknown speaker's identity. The figure 2.2 is the figure of the identification taxonomy. Speaker identification can be divided into closed-set iden-tification and open-set identification. For the closed-set identification, the task can divided into text-independent identification and text dependent identification, depend on the algorithm used for the identification. The goal of speaker recognition is to identify the speaker, independently of what the speaker is saying.

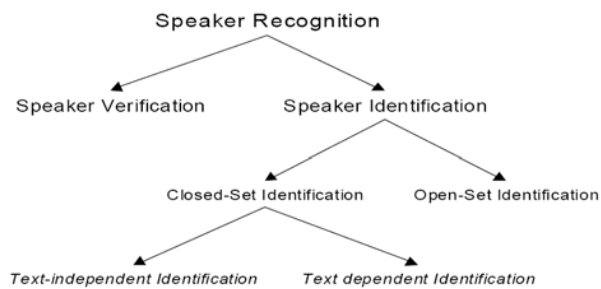

**Figure 1 Identification Taxonomy**

## 2.1 Speaker verification

Speaker verification is the process of accepting or rejecting the identity claim of a speaker. For the speaker verification, a speaker's recording is matched to previous recording, which was made during the voice registration period to produce the result. The result either is accepting or refusing the match between the speaker's recording and the previous recording.

## 2.2 Speaker Identification

Speaker identification is the process of determining which registered speaker provides a given utterance. Speaker identification required to make a claim on the identity of speaker from the Ns trained speaker in its user database. So, the role of automatic speaker identification (ASI) is more complex and the error of the system will increased if the Ns increased.

## 2.3 Close-set and Open set

Closed-set speaker identification occurs if all possible test utterances belong to one of the speakers that have been learned by the system. If a test utterance may be originating by a person that has not been shown to the system before, this is open-set speaker identification.

## 2.3 Text-dependent and Text Independent

Speaker recognition methods can be divided into two methods. There are text-dependent and text-independent. During the enrollment phase, the speaker's voice is recorded and typically a number of features are extracted. If the utterances used in the operational phase must be same with the utterances during the enrollment phase, this is text-dependent. On the other hand, for the text-independent, the utterances use in the operational phase is different with the utterances during the enrollment phase.

## 3 SPEAKER IDENTIFICATION SYSTEM

There are two phases in the speaker identification: enrollment phase and identification phase.

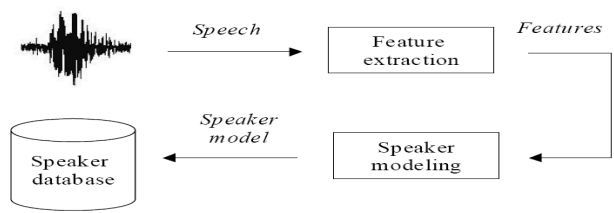

**Figure 2 Enrollment Phase**

During the enrollment phase, speech samples are collected from the speaker and convert the speech signal into set of feature vectors, which characterize the properties of speech that can separate different speakers. The collections of the enrollment

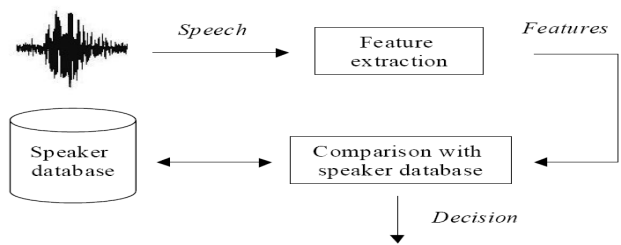

**Figure 3 Identification Phase**

In the identification phase, the test sample of unknown speakers is analyse and compared with all models from the speaker database. Both enrollment and identification phase has the same first step, feature extraction which converts the sampled speech signal into set of feature vectors, which characterize the properties of speech that can separate different speakers. The main purpose of this step is to reduce the amount of test data while retaining speaker discriminative information.

## 4 RELATED WORKS

### *Speaker Identification Based on Subtractive Clustering Algorithm with Estimating Number of Clusters*

Subtractive clustering algorithm which does the clustering without the initial guesses of location of the cluster centers and priori information about the number of clusters. The number of clusters is obtained by investigating the mutual relationship between clusters.
This algorithm will solve the problems that contain in the K-means and fuzzy c-means (FCM). The major problems are the improper initial guesses of cluster centers may degrade the performance and the number of clusters cannot always be defined a priori.
The new algorithm is based on subtractive clustering algorithm [11, 12] and mutual relationship of clusters [13]. First, cluster centers are obtained incrementally from adding one cluster center at a time through the subtractive clustering algorithm. Second, investigate the mutual rela-tionship of a cluster with respect to all the other clusters, to obtain the estimation of the number of cluster. Two clusters are statistically dependent is decide by the mutual relationship.

### Speaker Identification Using Reduce RBF Networks Array

The purpose of speaker identification is to determine a speaker's identity from the speech utterances. Essentially, speaker identification is a pattern recognition problem of a speech signal[16]. Both of the training and recognition processes are important for the speaker identification. Those processes are include the identification of discriminating features representing the specific characteristics of the speakers and the choice of the classifier. Radial basis function (RBF) networks choose because it successful use of GMM in speaker identification. The RBF network is a three-layer NN, which has the same underlying structure as the Gaussian Mixture Models (GMM) when Gaussian function is selected as the type of basis function in the RBF network[17]. Another reason for choose the RBF is RBF network easy to use. The key point in design of radial basis function networks is to specify the number and the locations of the centers. Input training data (IC) and both the input and output data (IOC) are the ways to obtained the vectors from the clustering algorithm. In this paper, IOC is choose because the structure of centers designed in the IOC is more effective. To select the suitable number or network centers, this method use the recursive orthogonal least-squares (ROLS) algorithm after the training process. Mel's frequency cepstral coefficients (MFCC) are extracted as the features for speaker characteristics. The system is composed of some binary classifiers, while the binary partitioned approach has been shown as an efficient solution for reducing training time[18].

## 5 OPERATION FRAMEWORK

In this paper, we focus on the performance of the speaker identification system using the clustering algorithm. In this case, we have applied the subtractive clustering as the clustering algorithm and RBF as the classification in speaker identification.

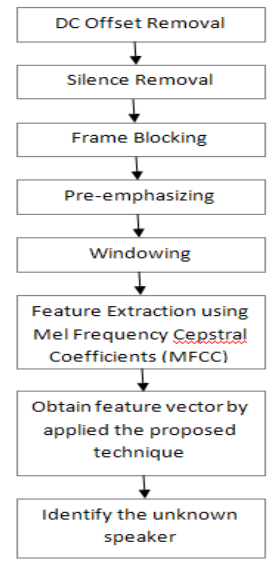

**Figure 4 Speaker Identification Framework**

### 5.1 DC offset Removal

Speech data are discrete-time speech signal, it often carry some redundant constant offset called DC offset [20]. These DC offset will effect quality of the information extracted from the speech signal. Consequently, we calculating the average value of the speech signal and subtracting this from itself[2].

### 5.2 Silence Removal

This process is to remove the silence periods from the speech containing silence frames to make the signal becomes more compact. The figure 3.4 is the example of the before and after the silence removal process. Silence frames are audio frames of background noise with a low energy level with respect to voice segments[2]. The signal energy in each speech frame is evaluated by equation (1).

$$E_i = \sqrt{\sum_{k=1}^{M} x_i(k)^2} \qquad i = 1, \ldots, N \qquad (1)$$

Where M is the number of samples in a speech frame and N is the total number of speech frames. Threshold is successively performed to detect silence frames with a threshold level determined by equation.

Threshold = Emin + 0.1 (Emax – Emin)

Emax – Emin are the lowest and greatest values of the N segment respectively.

### 5.3 Frame Blocking

The analysis of the speech signal is based on short-term spectral analysis. So, in order to analyze the speech signal, the speech signal is require to divided into fixed-length short frames. Figure 3.5 show that the speech signal is divided into overlapping frames with N analysis frame length and the frame is shifted with M samples from the adjacent frame. If the shifting is small, then the MFCC spectral estimated from frame to frame will be very smooth. If there is no overlapping between adjacent frames, the speech signals will be totally lost and correlation between the result LPC spectral estimates of adjacent frames will contain a noisy component[9].

### 5.4 Pre-emphasizing

Pre-emphasis is a technique used in speech processing to enhance high frequencies of the signal[2]. The main purpose of pre-emphasizing is to spectrally flatten the speech signal that is to increase the relative energy of its high-frequency spectrum. Pre-emphasis is important because:
  i. The speech signal generally contains more speaker specific information in the higher frequencies[21].
  ii. When the speech signal energy decreases, the frequency increases.

This allows the feature extraction process to focus on all aspects of the speech signal[2].

### 5.5 Windowing

In this process, the frames will be applied a window function to minimize the signal discontinuities at the beginning and the end of the each frame. A windowing function is used on each frame to smooth the signal and make it more amendable for spectral analysis[2]. A windowing function is used on each frame to smooth the signal and make it more amendable for spectral analysis. Usually for the speech processing, the smoother functions, Hamming Window is used.

### 5.6 Feature Extraction using Mel Frequency Cepstral Coefficients (MFCC)

A speech parameterization must be performed to convert the raw speech waveform into features vectors. After the windowing function is apply to the frames, the transformation technique, Fourier transform will be applied to the frames. The propose of the Fourier transform is to performed the transformation between time domain and frequency domain. Discrete Fourier Transform (DFT) and Fast Fourier Transform (FFT) are efficient implementation of Fourier transform for discrete time signals sampled at regular intervals[2]. Mel-frequency warping is used to reduced the large number or the channel for the FFT. This process is called "binning".

This step gives the linear-spectral domain feature vector, $O\_b^\wedge I$ (t), for bin b of the sample waveform at time t. Following that, logarithm transformation is applied to obtain the log-spectral values[2]:

$$O_b^l(t) = \log(O_b(t))$$

MFCCs can be obtained from the log-spectral values using inverse FFT[7]:

$$O_i^c(t) = \sum_{b=1}^{B} O_b^l(t) \cos(i(b-0.5)\pi / N_f)$$

### 5.7 Obtain feature vector by applied the Subtractive-RBF

This process is to generate the speaker model for each of the sample speech in the training phase and store into the database. First, we used the subtractive clustering to choose the network centers. The RBF network is a three-layer NN, which has the same underlying structure as the Gaussian Mixture Models (GMM) when Gaussian function is selected as the type of basis function in the RBF network[17]. Each hidden unit is a Gaussian kernel function with output related to the distance between the input vectors and the centroid of the basis function. The output layer forms a linear combiner which calculates the weighted sum of the outputs of the hidden units, and is given by,

$$y = \sum_{i=1}^{M} w_i \emptyset_{ci}(x) + b$$

where $x \in R^n$ is the input vector, $y \in R$ is the output node, $w_i$ is the weight from the $i$th RBF center, $b$ is the output bias, and the Gaussian kernel functions are

$$\emptyset_{c_i}(x) = \exp(-\|x - c_i\|^2 / r_i)$$

Where $c_i \in R^n$ is the $i$th center, and $r_i$ are the function widths. Once the centers have been fixed, the optimal linear weights can be determined straightforwardly by using a linear least squares algorithm, or taking the pseudo inverse.

### 5.8 Identify the unknown speaker

An unknown speaker's voice is represented by a sequence of feature vector {x1, x2 ….xi), and compared with the speaker model from the database. To identify the unknown speaker, we measure the distortion distance of two vector sets based on minimizing the Euclidean distance. The equation for measuring the distortion distance is

$$\sqrt{(p_1 - q_1)^2 + (p_2 - q_2)^2 + \cdots + (p_n - q_n)^2} = \sqrt{\sum_{i=1}^{n}(p_i - q_i)^2}$$

The more lower the value, that two vector sets is more similar and the lowest distortion distance of the speaker is chosen to be identified as the unknown person.

## 6 DISCUSSION

There are many techniques in speaker recognition system. The most common and widely use technique in speaker recognition are Hidden Markov models, Gaussian mixture models, pattern models, pattern matching algorithm, neural network and so on. But there are still need improved for the pattern recognition. By improved statistical modeling may produce better performance for the speaker recognition system. A better performance of the speaker recognition for the aspect speed and accuracy will attracted more people choose to use speaker recognition as the input method and replace the old input method.

## 7 CONCLUSION

In this project, the proposed technique is capable to identify the unknown speaker by search the best model in the database. From the previous works, we feel that The proposed technique can achieve faster training speed and more accurate for the speaker recognition task.

**Dr. Ibrahim A. Albidewi** is an Assoc.Prof in the Faculty of Computing and Information Technology in King Abdulaziz University, Jeddah Saudi Arabia. His current research interest includes Graphic and Multimedia, Image Processing. Pattern Recognition and Speech Recognition.

**Mr. Yap Teck Ann** is currently purchasing Master degree in Faculty of Computer Science and Information System. He received the B.Sc. Degree in Universiti Teknologi Malaysia in 2009. His current research interest includes Graphic and Multimedia, Pattern Recognition and Speaker Recognition.